# The Structure and Incentives of a COVID related Emergency Wage Subsidy

Jules Linden[1,2], Cathal O'Donoghue [1], Denisa M. Sologon[2]

[1] National University Ireland Galway, Ireland

[2] Luxembourg Institute of Socio-Economic Research (LISER)

**Abstract**

During recent crisis, wage subsidies played a major role in sheltering firms and households from economic shocks. During COVID-19, most workers were affected and many liberal welfare states introduced new temporary wage subsidies to protected workers' earnings and employment across the income distribution (OECD, 2021). New wage subsidies however marked a departure from the structure of traditional income support payments and required reform. This paper uses simulated datasets to assess the structure and incentives of the Irish COVID-19 wage subsidy scheme (CWS) under five designs. We use a nowcasting approach to update 2017 microdata, producing a 'near-real time' picture of the labour market at the peak of the COVID-19 crisis. Using microsimulation modelling, we assess the impact of different designs on income replacement, work incentives and income inequality. We find that changes in design substantially affected the functioning of the scheme. Our findings suggest that pro-rata designs support middle earners more and flat-rate designs support low earners more. We find evidence for strong work disincentives under all designs, though flat-rate designs perform better. Disincentives are primarily driven by generous unemployment payments and changes in work-related costs. Matching the structure of the unemployment payment and the wage subsidy only partially addresses work disincentives. The impact of design on income inequality largely depends on the generosity of payments. Earnings related pro-rata designs were associated to higher market earnings inequality. The difference in inequality levels fall however once benefits, taxes and work-related costs are taking into consideration. In our discussion, we turn to transaction costs, the rationale for reform and reintegration of the emergency wage subsidy. We find some support for the claim that design changes were motivated by political considerations. We suggest that establishing permanent wage subsidies based on sectorial turnover rules could offer enhanced protection to middle- and high-earners during emergencies and reduce uncertainty, the need for costly reform, and the risk of politically motivated designs.

---

[1]The Authors are grateful to the Irish Health Research Board, Irish Research Council and the Luxembourg Fond National de la Recherche (FNR/AFR Individual/14614512) for funding this research.
Corresponding Author : Jules Linden (email: jules.linden@liser.lu)

# The Structure and Incentives of a COVID related Emergency Wage Subsidy

1. **Introduction**

During recent crisis, job retention schemes played a major role in protecting workers and firms (OECD, 2021). At the onset of the COVID-19 pandemic, governments restricted movement of people and commercial activity to limit infections and shelter the healthcare system. The restrictions caused a significant demand shock and the state stepped in to support affected workers and firms. At the peak of the COVID-19 crisis, one third of workers in OECD countries were supported by a wage subsidy or short-time working (STW) scheme (OECD, 2021). Many liberal welfare states introduced novel emergency wage subsidies (Eurofound, 2020). We study the design and functioning of a novel emergency wage subsidy introduced to a liberal welfare state during a general demand shock. We compare five different designs to assess the functioning of each design with regard to income replacement, work incentives and redistribution.

Wage subsidies are one instrument utilized by policy makers to support employment levels. Wage subsidies, similar to other in-work payments, bridge the gap between workers' productivity and labour costs. Generally, in-work benefits raise the returns to working for low-wage workers, without increasing the employers' labour cost (Immervoll & Pearson, 2009). For in-work benefits to support employment however, employment should be restricted by high reservation wages. Where employment levels are constraint by low labour demand, like during emergencies, wage subsidies may be more efficient (Burns et al, 2010). Wage subsidies are usually paid to employers and subsidize labour costs. During emergencies, wage subsidies partially replace labour costs if employers are unable to pay wages. Like any policy affecting the returns to working, in-work benefits and wage subsidies affect incentives and the income distribution. The direct influence of taxes and benefits on household incomes and work incentives has been studied extensively (Immervoll and O'Donoghue, 2003). To our knowledge, no study compares the designs of new emergency wage subsidies with regard to work incentives and the income distribution.

During crisis, wage subsidies function primarily to maintain existing employment and firms' productive capacity. In the past, wage subsidies have been proposed to address high payroll costs, unemployment traps and to offset distributional and incentive concerns of energy taxation (Immervoll & Pearson, 2009; Böhringer & Rutherford, 1997). The literature on job retention schemes during crisis predominantly focuses on short-time work schemes (Hijzen & Venn, 2011; Cahuc, 2019). An emerging literature studies the efficiency (Rosenberg, 2020) and medium-run implications (Mayhew and Anand, 2020) of wage subsidies during the COVID-19 pandemic. The largest literature on wage subsidies studies their functioning (Heintz and Bowles, 1996) and effectiveness in generating new employment in normal times (Jaenichen and Stephan, 2009; Katz, 1998). Other studies explore the role of wage subsidies as automatic stabilizers (Zubrickas, 2020). We extend the literature by assessing the functioning of a novel temporary wage subsidy introduced to a liberal welfare state under five designs.

Multiple papers find that wage subsidies cushion the impact of COVID-19 on households' finances (Brewer and Tasseva, 2020; Sologon et al., 2020b). In the UK, Brewer and Tasseva (2020) find the earnings subsidy to be the main mechanism protecting household incomes. Distinguish between the effects of automatic stabilizers and emergency response policies, Bronka, et al (2020) find that that the latter limits the reduction of average household disposable incomes and poverty impacts. In Italy, Figari and Fioro (2020) find that the degree to which individual and household incomes are maintained varies with their characteristics. Indeed, the literature on the economic implications of Covid-19 shows heterogeneous effects of Covid-19 across industries and occupations (Papanikolaou and Smidt, 2020; Beirne et al., 2020). Most studies find higher levels of vulnerability for workers with lower levels of education and precarious employment (Benzeval et al, 2020). The hardest hit sectors employ proportionally larger shares of low-skill, low-earnings workers and those on precarious contracts (Eurofound, 2020). Growing evidence suggests that the likelihood of re-employment for some workers may be adversely affected by changes in work practices and skill requirement, such as telework and IT skills (Campello et al., 2020). The heterogeneity of the shock and the importance of the wage subsidy in sheltering households amplifies the need to understand how design choices affect households differently.

In recent decades, Ireland experienced two significant shocks and demonstrated the ability to innovate and adjust schemes when necessary. Ireland, like other the liberal welfare states, introduced a novel wage subsidy during COVID-19. The Irish COVID-19 wage subsidy scheme (CWS) was reformed five times, providing an adequate case for studying the functioning of different designs. We examine the level of protection offered by each design along the earnings distribution. We then focus on work incentives of affected workers. Lastly, we review the scheme's impact on income inequality. Understanding whom the scheme directed resources to is particularly relevant given size of the scheme and its substantial costs.

To compare the functioning of CWS under five designs, we use microsimulation modelling to compute a range of indicators capturing the level of income maintenance, work incentives and distributional characteristics for each design. We compute compensation rates, showing the wage subsidy as share of disposable income, and relative replacement rates, showing out-of-work income replacement relative to earnings under CWS. A range of inequality indexes show the distributive impact of the total policy response and scheme. We find that the flat rate schemes protect household incomes of low-earners most. Earnings-related schemes provide more support to medium and high earners, but substantially worsen work incentives at the bottom of the distribution. Flat-rate schemes provided better work incentives than earnings-related schemes. Matching wage subsidy payments and unemployment payments improves work incentives, though substantial disincentives remain. The impact of design on income inequality is less clear and largely depends on the generosity of payments. Earnings related pro-rata designs were associated to higher market earnings inequality. The difference in inequality however shrinks or reverses once we consider benefits, taxes and work-related costs.

In the discussion, we turn to the issue of reintegrating emergency schemes. The timing and nature of reintegration of emergency schemes is a key question for policy makers (OECD, 2021). Reintegrating an emergency wage subsidy will require reform or phasing out. In designing reforms, policy makers will have to decide where to maintain support and for how long. Reforms are costly and the reformed scheme should be financially and politically sustainable, whilst adequately support firms and workers in need. We discuss policy makers' rationale for reform and major trade-offs they face in reintegrating the scheme. We suggest that frequent reform was largely motivated by political, rather than economic, considerations. Therefore, we propose that extending automatic stabilizers to provide sufficiently generous to support middle class workers could help reduce the need for reform, reduce uncertainty and transaction costs. In absence of adequate automatic stabilizers, policy makers will need to introduce novel schemes, increasing uncertainty, transaction costs and opening the system up to political abuse.

## 2. Functioning and Structure of the COVID-19 Wage Subsidy Scheme

Across OECD countries, the initial policy response to COVID-19 focused on protecting businesses and workers (OCED, 2020a). In Ireland, CWS supported business and workers by covering businesses' labour costs, maintaining employment and providing earnings replacement. CWS was first introduced under the Employer COVID Refund Scheme (ECRS) on the 15th of March 2020 and made accessible to eligible employers that kept their employees on payroll (Revenue, 2020a). The subsidy was received by the employer through their payroll system as a reimbursement for wages paid. It was open to firms in all sectors, but conditional on a decline in turnover. Employers could make additional payments to bring total pay to the average net weekly pay for January and February 2020. Effectively, CWS subsidised workers' wages and functioned as a benefit to workers, shifting the balance of between public and private sector expenditure. The subsidy was cumulative with a state pension and with the Working Family Payment. Where workers lost their employment, a generous Pandemic Unemployment Payment (PUP) provide income support. Households received additional support through mortgage interest payment deferrals and extended fuel allowances. Businesses were further supported through tax deferrals, notably suspended payroll tax, debt warehousing, grants and access to cheap loans.

CWS was reformed multiple times and it operated under three different names. Reforms concerned the structure of the scheme, i.e. whether it was paid as a flat-rate or pro-rata payment, its generosity at different points of the earnings distribution, and the condition under which it could be accessed. On March 15th, the scheme was administered as a flat-rate payment of €203 per worker per week, matching the unemployment payment. Under the transitional Temporary Wage Subsidy Scheme (TWSS), the scheme moved to a complex earnings-related pro-rata payment for lower incomes on March 26th, with a flat rate structure for workers median earners and a cap at high earnings (Revenue, 2020a). The operational TWSS introduced on May 4th increased payment levels for low earners and introduced a tapered approach at earnings higher than median. On September 1st, the Employer Wage Subsidy Scheme (EWSS) moved to a less generous earnings flat-rate structure and the turnover threshold was increased from 25% to 30% (Revenue, 2020b). In October, EWSS payments and the number of bands were increased. EWSS was open to new hires and seasonal workers. Appendices 1 and 4 summarize these changes.

March and May reforms sought to addressed concerns around work incentives and inadequate support for lower earners. September and October reforms shifted the focus to reintegration of the scheme and financial sustainability. In September, the emphasize was on supporting viable firms, improve work incentives and encouraging employment (Department of Finance, 2020). During a second wave of the virus in October, the scheme returned to generous levels.

Changes in the state of the crisis and policy makers priorities also reflect in PUP's design. PUP was reformed five times between March and October. PUP reforms preceded CWS reforms. Design changes concerned the schemes' generosity and the number of earnings bands. Reforms are summarized in Appendix 2. Appendix 3 provide a stylized comparison between CWS and PUP.

## 3. Theory: Policy aims and design

In the context of a major job loss, such as the Great Recession or the COVID-19 crisis, temporary wage subsidies are purposed to maintain employment and provide income support (ILO, 2020). In the past, wage subsidies were proposed to address a range of issues. Early proponents emphasized their stabilizing effects during economic downturns and benefits for international competitiveness (Kaldor, 1936). During the 1980s, wage subsidies were proposed to combat high levels of unemployment in developed countries by reducing the gap between labour cost and productivity resulting from high payroll costs (Adnett and Dawson, 1996). During the 1990s, wage subsidies were employed to address incentive distortions caused by traditional benefits and unemployment traps faced by the long-term unemployed and single parent households (Snowder, 1994). More recently, wage subsidies were discussed to mitigate the impact of energy taxation on international competitiveness, the income distribution and work incentives (Böhringer and Rutherford, 1997; Stiglitz, 2019).

During recent crisis, wage subsidies functioned primarily as job retention schemes, maintaining existing employment by covering a share of firms' labour costs. Job retention schemes were widely adopted during the COVID-19 crisis and the Great Recession, though their design varied substantially.[2] Some countries expanded existing short-time working (STW) schemes, while others introduced new temporary wage subsidies. Temporary wage subsidies were popular in liberal welfare states (such as Australia, the UK and Ireland), while most continental European welfare states (like France, Belgium and Germany) relied on pre-existing STW schemes (OECD, 2020). Cantillon et al (2021) suggest that the extent to which continental systems introduced new policies depends on the extent to which they moved towards an Anglo-Saxon model. In this liberal block, where unemployment payments are administered as a flat-rate allowance, STW payment levels (in line with unemployment payments) were insufficiently generous to support middle and higher earners affected by COVID-19 and governments opted for novel emergency wage subsidies. Such emergency wage subsidies typically involve direct monetary transfers to employers for workers retained on the payroll. Emergency wage subsidies help firms avoid the costly processes of separation, re-hiring, and training (Giupponi and Landais, 2018), maintaining firms' productive capacity and workers' income.

In designing wage subsidies, policy makers decide on targeting, coverage of the scheme and the conditionality under which it is accessed (Burns et al, 2010). Firstly, policy makers decide to pay the subsidy to employers or employees. Employee-based wage subsidies, usually targeted at individuals with low labour market attachment, help overcome unemployment traps and increase labour supply (Immervoll & Pearson, 2009). Employer-based wage subsidies reduce labour costs and increase labour demand, and are often targeted at firms in specific sectors (Adnett and Dawson, 1996). During emergencies, employer-based wage subsidies, such as CWS in Ireland or 'furlough' in the UK, function as a benefit to the employer and the employee. They reimburse employers for employee's wages, allowing employers to meet their liquidity needs and to bridge periods of reduced consumer demand without resorting to layoffs. Employees retain their job and have their earnings maintained. Secondly, policy makers must decide to subsidize jobs or hours. STW schemes compensate workers for hours not worked, allowing firms to unilaterally decrease hours worked, with the state compensating workers' forgone hours at the rate of unemployment payments[3]. Emergency wage subsidies in contrast are generally paid to the employer, covering firms' labour costs irrespective of hours worked. Wage subsidies therefore provide a stronger incentive for firms to restart production and may facilitate a swifter recovery (OCED, 2020a). Simultaneously however, wage subsidies prevent workers from relocating to more productive firms and may lead to higher deadweight losses.

---

[2] OECD (2020a) provides an overview of selected job retention schemes during COVID-19.
[3] Studies frequently point towards positive effects of STW on employment along the extensive margin (worker head count) but negative effects along the intensive margin (hours per worker) (Giupponi and Landais, 2018).

Deadweight loss is often cited as major concern with wage subsidies. Deadweight losses occur when a subsidy supports workers that would have been retained in the absence of the scheme (Adnett and Dawson, 1996). Targeting and conditionality help address concerns over deadweight losses and excessive fiscal costs (Adnett and Dawson, 1996). If program costs or deadweight losses are a concern, policy makers can cap the subsidy at higher earnings levels or introduce tapers. This makes the scheme more complicated and increases administrative costs, but likely reduces deadweight losses and overall program cost. During COVID-19, the large emphasize on speed and solidarity meant that concerns about fiscal costs and deadweight losses were secondary, and most initial schemes were universal in targeting (OECD, 2021). In most countries, the subsidy was conditional on reductions in turnover or sales, irrespective of firm size or industry. An emphasize on protecting workers and firms was seen as socially optimal, especially with regard to an expected v-shaped recovery.

Broad access to the scheme meant that recipients would likely differ in their previous wage and need for earnings replacement. Policy makers had to decide on the generosity of the scheme, and whom to direct resources too. The generosity and structure of the scheme affects the level of protection offered to different jobs. Flat-rate structures provide some support to all workers but supports low-earning jobs more. Earnings-related subsidies improve the incentives to retain high-wage workers (Heintz and Bowles, 1996). The choice of structure will also affect workers' incentives. A divergence in wage subsidy's and the unemployment benefit's structure may cause a type of unemployment trap, whereby workers are better off unemployed, particularly when wage subsidies only cover a share of previous earnings and employers are unable to top-up the wage subsidy payment. Work disincentives may be amplified by work-related costs and caring duties. Firms may consequently not take-up the subsidy if workers that are better off unemployed succeed in putting sufficient pressure on firms to lay them off. Conversely, firms may struggle to (re-)hire workers if workers benefit from generous unemployment payments. Incentive issues are of particular concern where multiple emergency schemes were introduced, or existing out-of-work payments enhanced.

Firms' decision to take-up is also affected by the scheme's transaction costs and administrative burdens. Firms will evaluate the generosity of the scheme relative to its transaction costs and the cost of firing and hiring. Complex schemes with significant information requirements come at higher cost (Currie, 2009). Administrative burdens and the associated transaction costs arise from information obligations requiring firms to collect data and from efforts in understanding the scheme's functioning and training of employees (Bennett et al, 2009). They grow with the complexity of the scheme and with frequent reforms. Simpler schemes may be prefered if high take-up is essential for the success of the scheme and programme costs are of lesser concern.

A last design choice concerns the duration of the subsidy. A pertaining issue with emergency schemes and wage subsides is to distinguishing viable firms in need of temporary assistance from those in decline (Adnett and Dawson, 1996). During COVID-19, access to wage subsidies was frequently conditional on reductions in sales or turnover (OECD, 2020). The duration of the subsidy was temporary, though not set in stone, and eligibility was re-evaluated on a rolling basis. Re-evaluating eligibility based on turnover levels likely reduces the scheme's deadweight loss and costs, but can also put stress on businesses experiencing seasonal fluctuations in sales, generate uncertainty and provide incentives for businesses to restrict production (Bundell, 2002). In volatile environments, the decision to phase out the scheme is difficult. If the scheme is phased out too early, viable businesses are lost and policy makers face political opposition. Keeping subsidy in place for too long supports unviable businesses, prevent workers from reallocating effectively and causes excessive fiscal costs.

During emergencies, policy makers can introduce new wage subsidies to protect firms, incomes, and jobs. In designing emergency wage subsidies, policy makers redistribute resources. They decide on which jobs to protect and how generous the level of earnings protection should be. These decisions affect the scheme's design, its complexity, cost, and redistributive capacity. The scale of emergency wage subsidies during recent crisis raises the importance of understanding how design choices affect the level of earnings support, whose earnings they support and how they affect work incentives.

## 4. Methodology

New in-work schemes affect work incentives and household incomes. The interactions of a new wage subsidy with the remainder of the tax benefit system requires a methodology that can accommodate heterogeneous households, simulate diverse policies and changes in labour market circumstances. Microsimulation modelling fulfils this requirement and allows to study new policies in relation to the existing system. Using microsimulation, we compute household disposable incomes. To account for the specific expenditure changes related to COVID-19, we adjust our disposable income context for housing costs, work-related expenditure (child care and commuting) and capital loss (O'Donoghue et al, 2020). The fast-paced development of the crisis means that microdata usually used for microsimulation no longer reflects the current labour market. We update the underlying microdata by calibrating to the most recent available macroeconomic data reflecting the labour market at the peak of the COVID-19 crisis. Before describing the nowcasting-microsimulation approach, we present three key indicators used in the analysis of the emergency wage subsidy's functioning.

*4.1. Compensation rate*

Before utilizing microsimulation, we provide a stylized view of the level of earnings replacement provided by the different CWS designs. The contribution rate (CR) shows the CWS payment as share of pre-crisis gross earnings.

$$CR = \frac{CWS\ payment}{Pre-crisis\ Gross\ Earnings}$$

In computing the average CR, we weight CR according to the share of workers situated in a given income range. In computing CR for different deciles, we give equal weights to all income ranges in a given decile. Thereby, we assume that the likelihood to receive CWS is equal across deciles and income ranges.

*4.2. Net replacement rate*

To understand the impact of new instruments, we need to consider how they interact with pre-existing elements of the tax-benefit system. The net replacement rate shows the level of income stabilization relative to a baseline level (Immervoll and O'Donoghue, 2004). Here, we show the share of equivalized adjusted disposable income covered by CWS for recipients.

$$RR_{net} = \frac{CWS\ Payment}{Equivalized\ adjusted\ Household\ Disposable\ Income}$$

Adjusted household disposable income consists of the sum of pre-tax incomes (market income, capital income, private pension, rent income, investment income) and the difference between public transfers and taxes (social security contributions, income tax, working family supplement, child transfer, PUP and CWS). In modelling market earnings of CWS recipients, we assume that employers top up recipient's wages, so that recipient's gross income is fully maintained[4]. The public health response to COVID-19 introduced important changes in consumption patterns, expenditure and returns to investments. We extend our disposable income concept to account for changes in investment income, housing costs and work-related costs (e.g. commuting and childcare). We account for household composition by equivalizing disposable incomes using the squared adult equivalence scale.

*4.3. Relative replacement rates*

The introduction of a generous unemployment payment changed the returns to working. CWS aims to maintain employment, meaning that work incentives are central to the scheme's success. The relative replacement rate shows adjusted disposable income under CWS relative to adjusted disposable income under PUP for recipient households, capturing the financial incentive to stay in work.

The replacement rate is found by taking:

$$RR_{relative} = \frac{Out-of-Work\ Adjusted\ Disposable\ Income\ under\ PUP}{In-Work\ Adjusted\ Disposable\ Income\ under\ CWS}$$

In order to capture the actual returns to staying in work, we utilize simulation-based replacement rates. We estimate in-work and out-of-work disposable incomes on a sample of households in receipt of CWS.

*4.4. Nowcasting*

The ideal approach for modelling the impact of CWS involves using household survey data comprising extensive information on household incomes. The sudden impact and swift development of the crisis however confronts us to a lack of up-to-date survey data. The main survey with data on household incomes is the Survey of Income and Living Conditions, last release in 2018 with incomes referring to 2017.

We use a "nowcasting" methodology (O'Donoghue and Loughrey, 2014) to update the latest available survey data using recent data on employment and prices. By calibrating a microsimulation model of household incomes, taxes and benefits to more recent aggregate statistics, we produce a near-real time picture of the population and those affected (O'Donoghue, 2014; Atkinson et al, 2002). By doing so, we follow recent developments in the field (see O'Donoghue et al. 2020; Sologon et al. 2020b), going beyond existing approaches that apply price inflation factors and proportional changes in the employment rate in specific industries (Navicke et al., 2014). Datasets containing more recent aggregate statistics are: (i) the Labour Force Survey, available on a quarterly basis at a 6-week lag; (ii) the Live Register data and (iii) price data, available on a monthly basis.

To adjust our income distribution data to macro-economic changes, we utilise a dynamic ageing mechanism. We update the data by estimating a system of equations that model the income generating process utilising a dynamic modelling approach (Li and O'Donoghue, 2014). We use the generic household income-generation model (IGM) developed by Sologon et al. (2020a), which simulates the labour market and household market income distribution as a function of personal and household attributes and generates counterfactual distributions under alternative scenarios. Taxes and benefits are simulated using the NUI Galway microsimulation model developed for studying the impacts of an economic crisis (O'Donoghue et al., 2018).[5]

---

[4] In May, an average of 86% of recipients received a top-up (Revenue, 2021).
https://www.revenue.ie/en/corporate/documents/statistics/registrations/wage-subsidy-scheme-statistics-07-may-2020.pdf

Fundamentally, the nowcasting model is composed of:
- an income generation model, that describes the distribution of household market income;
- a tax-benefit model, that converts the distribution of market incomes into the distribution of disposable income;
- a calibration component, that calibrate the simulation from the income generation model to external statistics.

The model diagram is illustrated in Figure 1.

*4.5. Income generation model*

The income generation model involves estimating a set of equations describing the generation of the main components of household disposable income; labour (from employment or self-employment), capital (investment and property) and other non-labour income.

Disposable income, $Y_{D,t}$, at time t depends upon market income $Y_{M,t}$ (labour income, capital and other income), benefits $B(Y_{M,t}, Z_t, \theta_t^B)$ and taxation $T(Y_{M,t}, Z_t, \theta_t^T)$, which in turn depend on personal and family characteristics, $Z$ and tax-benefit parameters $\theta$. Given the nature of the shock, and the multi-faceted impact on household living standards, it is necessary to utilise an augmented version of disposable income ($Y^*$). We adjust disposable income for (i) work-related expenditures $C_t$; (ii) housing costs $H_t$; (iii) capital losses $Q_t$.

$$Y_{D,t} = Y_{M,t} - T(Y_{M,t}, Z_t, \theta_t^T) + B(Y_{M,t}, Z_t, \theta_t^B) - H_t - Q_t - C_t.$$

The IGM (see Sologon et al., 2020a) relies on the estimation of a system of hierarchically structured multiple equations for the components income, that describe the presence of the market income source $(I_{i,t})$ and its level $(Y_{i,t})$:

$$Y_{M,t} = \sum_{i=1...m} Y_{i,t}^* = \sum_{i=1...m} \left\{ Y_{i,t}(Z_{i,t}^Y, \theta_{i,t}^Y, \varepsilon_{i,t}^Y) \times I_{i,t}(Z_{i,t}^I, \theta_{i,t}^I, \varepsilon_{i,t}^I) \right\}$$

Thus, for all income sources we apply a two-step procedure: first, we model the presence of the income source and second their level, conditional on receiving the income source.

The presence of the income source is modelled in the *labour market module*. For labour income, we first estimate the probability of being in work using logistic and multinomial models. For those in work, we estimate the probability of being employed versus self-employed, their occupation, industry, sector, type of contract. For those out of work, we model the probability of being unemployed, retired and inactive. For the remaining income components, we estimate the probability of having a given income source. The level of the income is modelled in the *income module*. Here we use log-linear models as a function of demographic and labour market characteristics, conditional on labour market status.

---
[5] This methodology was applied for historical analysis (Sologon et al., 2019).

At this stage, we save the parameter estimates and the individual specific errors from all models ($\varepsilon_i$). These parameter estimates describe the empirical associations between the vector of personal characteristics and the income components, whereas the vector of residuals connect the model predictions with the income sources. We subsequently use these projections to simulate counterfactual distributions of market incomes. To use the estimated probabilities from logistic models within a Monte Carlo simulation, we draw a set of random numbers such that we predict the actual dependent variable in the estimation data (see Sologon et al. (2020a) for the method). The disturbance terms are normally distributed, recovered directly from the estimation data for those with observed incomes, or generated stochastically for those without a specific income source in the original data.

We use the tax-benefit model for Ireland described in O'Donoghue et al. (2018) to convert market incomes into disposable incomes by calculating taxes and benefits:
$$T(Y_{M,t}, Z_t, \theta_t^T); B(Y_{M,t}, Z_t, \theta_t^B)$$

Overall, we can summarize the income generation model by the parametric structure:
$$Y_D = m(Z, LM(\theta), R(\theta, \tau), TB(\theta))$$

where $LM(\theta)$ represents the labour market structure, $R(\theta)$ represents the returns and price structure and $TB(\theta)$ represents the tax-benefit parameters. $\tau$ is the indexation system applied to align the monetary variables with the tax-benefit parameters to avoid fiscal drag.

Figure 1. Model Diagram

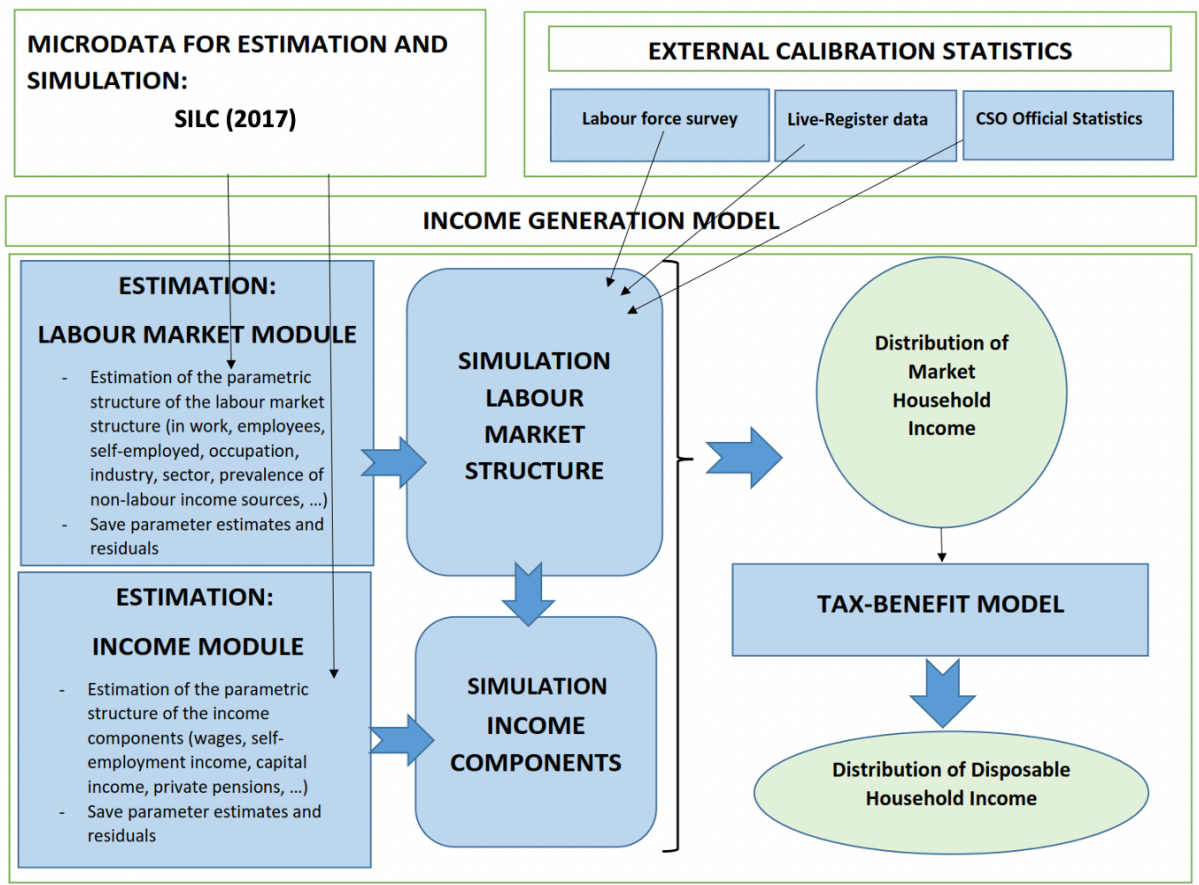

*4.6. Calibration component*

Our aim is to use the IGM to simulate the impact of an economic shock and its associated policy response by starting from the latest available survey data and nowcasting the data to reflect the macro-economic climate in Ireland during COVID-19. The simulations involve calibrating the econometrically estimated equations in the IGM to more timely external control totals, reflecting the labour market structure, income growth and price changes. The calibration mechanism or alignment is drawn from the dynamic microsimulation literature and described in Li et al (2014).

The objective of calibrating a microsimulation model is to ensure that the simulated output matches exogenous totals. We utilise three types of alignment: (i) binary discrete data, (ii) discrete data with more than two choices and (iii) continuous data. In short, we build up our microdata from the time of collection (period s) to the present and we calibrate the IGM data to external control totals reflected in the macro trends. This will ensure that the IGM is describing the targeted period t (during the crisis).

The simulations are done in 3 steps. First, we perform a labour market transformation by re-simulating the labour market module (see Figure 1) and aligning the simulations based on the evolution of the labour market regarding in-work shares by age and gender, the employment shares by industry, occupation and gender, the unemployment shares by gender. Each individual receives an updated labour market status, aligned to period t: $LM_t(\theta)$.

Next, we perform the returns and price transformation. First, the changes in labour market characteristics result in new simulated values for income components. Second, the monetary variables are uprated to reflect the evolution in average wages and price. Earning are indexed to reflect the differential income growth by industry and occupation. For those with capital income, we assign the probability of holding shares across the age-income distribution on the basis of Monte Carlo estimates using Iterative Proportional Fitting (IPF) and we simulate an average change in the capital value or capital loss at the median (Wong, 1992). Each individual receives an updated vector of income components, aligned to period t: $R_t(\theta, \tau)$. Lastly, we update the tax-benefit parameters to reflect the target period: $TB_t(\theta)$.

In terms of work-related expenditures, we model and simulate commuting costs and childcare costs. For commuting costs, we first estimate the probability of commuting by car or by public transport as a function of occupation, industry, education, location, and age group. Second, we estimate models for both public transport and motor fuels as a function of household characteristics, disposable income, social group and number of workers and predicted the proportional increase in costs as a function of the number of workers in a household relative to members not working. We assume a flat commuting costs across households, adjusted for the age.

The distribution of childcare costs per week, by family type and disposable income decile, is approximated using IPF. These averages are, in turn, used to calibrate the simulations based on the estimated models for having childcare and level of childcare expenditure (integrated in IGM). We obtain the nowcasted distribution of household disposable income that reflect the situation at time t:

$$Y_{t,D}^* = m\big(Z, LM_t(\theta), R_t(\theta, \tau), TB_t(\theta)\big).$$

We assess the impact of COVID-19 wage subsidy and other COVID-10 policies on the base 2020 income distribution by comparing the counterfactual distribution $Y_{t,D}^{**}$ under alternative wage subsidy policies. Comparing counterfactual distributions allows us to vary the design of specific policies and evaluate how different designs affect income replacement, work incentives and the income distribution under different labour market conditions.

## 5. Data

In order to utilize household survey data in a rapidly evolving context, we calibrate household survey data to up-to-date external macro control totals. This requires two types of data; (i) microlevel household survey data to perform tax benefit calculations and estimate observed relationships in the data; and (ii) Macrolevel calibration control totals to align the outdated microdata to recent changes in the labour market.

*5.1. Microdata*

The micro data stems from the 2017 Survey of Income and Living Conditions (SILC). SILC provides us with the information required for tax-benefit modelling and for estimating the relevant labour market and income parameters. It contains information on incomes, living conditions, labour market characteristics and demographics. SILC relies partially on survey data and partially on register data. Approximately 80% of respondents supplied their national social security number to assess their administrative data in relation to their benefit entitlement (Callan et al, 2010). Presentiveness of the dataset on the national level is achieved with respect to age, gender, household composition and region by utilizing a national weighting mechanism.

Time mismatch in measuring personal characteristics and income is a well-known issue in using SILC for tax-benefit modelling. The dynamic aging methodology applied in this paper helps us address the time mismatch bringing survey data to quasi real-time by calibrating them to external control totals (O'Donoghue et al., 2014).

*5.2. Calibration data*

In periods of high economic volatility, real-time modelling requires data that closely reflects the current period of the crisis. SILC data is calibrated using Labour Force Survey (LFS) and Live register data to ensure that the microdata reflects the macroeconomic climate during the COVID-19 pandemic. We adjust SILC data to the current employment situation, the provision of CWS and PUP and requests for mortgage deferral.

LFS provides us with the external calibration totals of the share of workers in work by age and gender, and employment shares by industry, occupation, and gender, as well as unemployment shares by gender in May 2020, at peak unemployment. Live Register provides data on the take-up of the COVID-19 Special Illness Benefit, CWS and PUP by industry. We model the change in employment for May 2020 by subtracting the number of PUP, normal unemployment benefit and CEIB recipients from February 2020 employment levels. LFS data is available quarterly with a 6-week lag and Live Register data is published weekly. Together, LFS and Live register data enable us to understand the characteristics of households and "update" their incomes to near real-time using estimation on the 2018 SILC dataset.

*5.3. Official Statistics and disposable income adjustment*

We propose an adjusted version of disposable income that accounts for changes in expenditure and capital returns throughout the crisis. Data sources include the 2016 Household budget survey and the Irish Household Finance and Consumption Survey. The average modelled commuting costs was of €9.17 per worker per week, €14.42 for two workers and €23.82 for three workers. Regarding capital returns, we utilize IPF to create an approximation of the distribution of share values across age and income. Changes in capital returns follow observed changes in the ISEQ index. The adjusted disposable income concept is described in more detail in O'Donoghue et al (2020).

## 6. Results

*6.1. Generosity*

The scheme's generosity, and whom it was directed to, changed with subsequent reforms. Figure 2 describes the stylized structure of the wage subsidy under different designs. The initial flat-rate ECRS was implemented swiftly and matched the existing unemployment payment in structure and generosity. On March 26th, the transitional TWSS moved to an earnings-related pro-rata structure, leading to a mismatch in unemployment and wage subsidy payments (see Appendix 3). The reform shifted the scheme's generosity towards the middle of the earnings distribution and reduced payment levels for low earners. The operational TWSS increased the generosity for the lowest earning 25% of workers[6], and introduced a tapered approach for higher earners based on the wage paid by the employer. September's EWSS reform marked a change in emphasize from protection to reintegration. The two-band flat rate EWSS (September) reduced generosity for close to all workers, with the highest payment level matching pre-crisis unemployment payments. The introduction of additional bands under EWSS (October) increased generosity for all eligible workers. Both EWSS designs excluded the lowest earning 4% and the highest earning 10%.

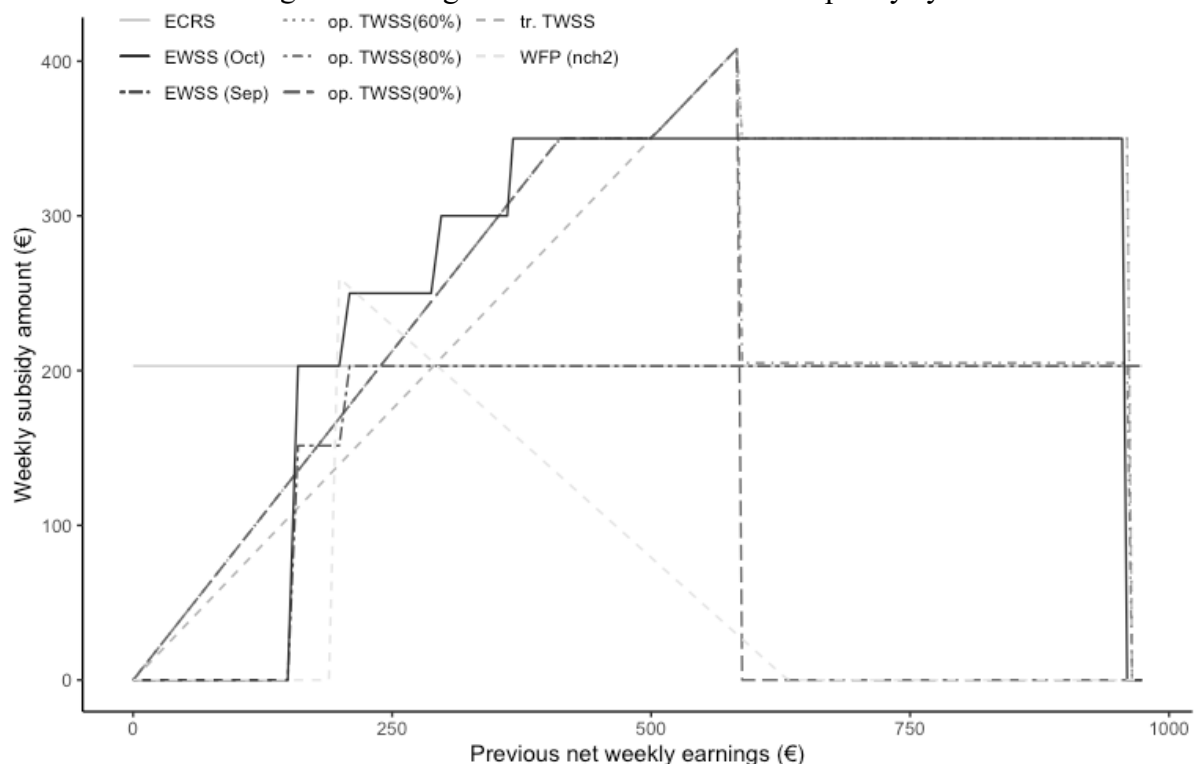

Figure 2. Budget constraint under different policy systems

Note: tr. TWSS = transitional TWSS; op. TWSS (60%) = operational TWSS with employer paying 60% of

---

[6] https://www.cso.ie/en/releasesandpublications/ep/-peaads/earningsanalysisusingadministrativedatasources2018/distribution/

previous net earnings; op. TWSS (80%) = operational TWSS with employer paying 80% of previous net earnings; op. TWSS (90%) = operational TWSS with employer paying 90% of previous net earnings; WFP (nch2) = Working Family Payment with two children.

Table 1 shows the level of earnings replacement offered by each design at the bottom, middle and top of the earnings distribution. The simple flat-rate ECRS and EWSS (Sep) schemes provided high levels of earnings replacement for low earners, but provide limited support to middle and high earners. Earnings-related pro-rata schemes replace middle and high incomes at higher rates relative to simple flat-rate designs. EWSS (Oct) earnings-related flat-rate multi-band design provides generous support across the earnings distribution, suggesting that flat-rate designs can provide levels of support similar to pro-rata schemes across the income distribution if payments vary by earnings band and the number of bands is sufficient.

Table 1. CWS compensation rate by decile (as share of previous gross pay)

| Decile | ECRS | trTWW | opTWSS | EWSS(Sep) | EWSS(Oct) |
|---|---|---|---|---|---|
| Bottom | 1 | 0.70 | 0.85 | 0.87 | 1 |
| Median | 0.37 | 0.60 | 0.64 | 0.37 | 0.64 |
| Top | 0.14 | 0.27 | 0.27 | 0.14 | 0.25 |
| Average | 0.29 | 0.43 | 0.46 | 0.28 | 0.45 |

Note: Deciles based on eligible earnings range. Average weighted in relation to workers in decile.

Table 2 provides an overview of net replacement rates of recipient households for selected deciles, providing a more nuanced view of the level of income protection. The simulated results largely reflect the stylized results presented in Table 1. A notable difference lies in the level of income replacement offered to higher incomes. Once we compare the CWS payment to adjusted disposable income, the levels of income replacement increase relative to the stylized results.

Table 2. Net replacement rate ($RR_{net}$) by decile (in %)

| Decile | ECRS | tr. TWSS | op. TWSS (May) | EWSS (Sep) | EWSS (Oct) |
|---|---|---|---|---|---|
| Bottom | 116.0 | 106.1 | 99.4 | 89.2 | 110.1 |
| 3rd | 78.0 | 66.9 | 76.2 | 63.6 | 89.9 |
| Median | 62.6 | 55.8 | 67.7 | 56.2 | 79.3 |
| 7th | 52.6 | 41.8 | 58.2 | 51.5 | 73.1 |
| Top | 52.8 | 46.1 | 59.2 | 56.7 | 76.3 |
| Average | 58.2 | 49.7 | 64.6 | 55.4 | 77.8 |

Note: Net replacement rates are computed based on household adjusted disposable income equivalized using the squared adult equivalent scale.

*6.2. Work incentives*

The primary purpose of CWS was to maintain employment. Businesses may struggle to retain workers if workers benefit little from taking up work. Throughout the pandemic, PUP supported workers that lost their employment, and was workers' main alternative to CWS. The relative replacement rate compares the adjusted household disposable income under both schemes, CWS and PUP.

In evaluating the incentive to remain employed under CWS, we are primarily interested in the share of workers with high RRs. Table 3 shows the share of workers that would be financially better off if they claimed PUP than if they remained employed under CWS. Under all COVID-19 systems, the financial incentive to move into unemployment was strong for a substantial number of workers. This is primarily driven by a generous flat-rate PUP. The flat rate ECRS and EWSS provided least disincentives to work. Earnings-related pro-rata systems show larger disincentive effects. This can be explained by a mismatch in payment structure between PUP and CWS. The mismatch in payment structure does however not fully explain disincentives once other policies and work-related costs are considered. Disincentives remain strong for a substantial share of workers under EWSS (Oct). Under this system, PUP and CWS were matched in structure and payment levels.

Table 3. Share of recipients by relative replacement rate ($RR_{relative}$) band

| RR band | ECRS | tr. TWSS | op. TWSS (May) | op. TWSS (Jul) | EWSS (Sep) | EWSS (Oct) |
|---|---|---|---|---|---|---|
| 70-89 | 26.4 | 33.2 | 31.8 | 32.1 | 30.7 | 34.8 |
| 90-99 | 7.9 | 14.0 | 17.3 | 16.2 | 11.2 | 15.1 |
| >=100 | 13.0 | 25.0 | 22.4 | 22.7 | 14.6 | 12.3 |
| Total | 47.3 | 72.2 | 71.5 | 71.0 | 56.5 | 62.2 |

Note : ECRS = Flat-Rate CWS and Low Flat-Rate PUP; tr. TWSS = Earnings-Rate CWS and High Flat-Rate PUP; op. TWSS(May) = Earnings-Rate CWS and High Flat-Rate PUP; op. TWSS(Jul) = Earnings-Rate CWS and Two-Rate PUP; EWSS(Sep) = Flat-Rate CWS and Three-Rate PUP ; EWSS(Oct) = Flat-Rate CWS and Four-Rate PUP).

Operational TWSS (Jul) not shown in previous results, represents the policy system in July 2020. Under this system, CWS' structure did not change but PUP was made earnings related, introducing two bands, with the lower band paying less than previous. Relative to the operational TWSS system of May, the share of workers facing high levels of work incentives only fell marginally, suggesting that disincentives are not only affected by high levels of PUP, but that work-related costs play an important role.

*6.3. Incidence and distributive impact*

In this section, we shed further light on the distributive impact of COVID-19 and its policy response. CWS was open to all workers that previously earned less than €1462 gross weekly. Table 4 shows a series of indexes describing the impact of the COVID-19 shock on household income inequality and the role of CWS in mitigating its impact. The Gini in market income excludes the wage subsidy payment. Where workers received CWS, their market income was computed as workers earnings minus the CWS payment. Labour market inequality rose substantially during the COVID-19 crisis, reach close to 0.60 under all systems up from 0.45 pre-crisis, illustrating the significant heterogeneity of the shock across workers. Under the two less generous flat-rate designs (ECRS and EWSS (Sep)), market income inequality was lowest, as higher earners experienced large reductions in earnings. Notably, under the generous flat-rate EWSS (Oct) design, labour market inequality was higher than under other flat-rate designs and under the less generous pro-rata transitional TWSS design. Gini in adjusted disposable incomes are lowest where payments of CWS and PUP were highest, suggesting that income inequality is primarily affected by the generosity of payments, rather than their structure.

Table 4. Progressivity and redistribution of CWS on equivalized household disposable income

|     | Gini | ECRS | tr. TWSS | op. TWSS (May) | EWSS (Sep) | EWSS (Oct) |
|-----|------|------|----------|----------------|------------|------------|
| (1) | Gini in Market income (excl. CWS) | 59.2 | 60.0 | 60.8 | 59.4 | 60.3 |
| (2) | Gini in Gross income ((1) + benefits, incl. CWS & PUP) | 36.9 | 35.8 | 35.3 | 36.4 | 35.5 |
| (3) | Gini in Adjusted disposable income ((2) – taxes – work related costs) | 31.8 | 31.1 | 30.8 | 31.9 | 31.0 |
| (4) | Gini in Adjusted disposable income without CWS ((3) – CWS) | 34.3 | 33.2 | 34.2 | 34.0 | 34.8 |
| (5) | Benefit redistribution (RS) ((4) – (3)) | 2.5 | 2.1 | 3.4 | 2.1 | 3.8 |
| (6) | Benefit Regressivity (K) | 0.63 | 0.65 | 0.65 | 0.66 | 0.67 |

Note : Computed based on a sample of households with (at least) one member in receipt of CWS. Household adjusted disposable income equivalized using the squared adult equivalence scale. K = Kakwani index; RS = Reynolds Smolensky index.

CWS' level of benefit redistribution, as measured by the Reynolds-Smolensky (RS) index, was highest under generous designs (op. TWSS and EWSS (Oct)). The RS index shows the extent to which CWS cushions the impact on income inequality and thus its redistributive effect. Additionally, we present the Kakwani index (Kakwani, 1977), which captures CWS' progressivity/regressivity. The Kakwani index indicates that the overall policy reponse was regressive in structure, paying more for higher earners. CWS was regressive under all designs, though this was less pronounced for the simple flat-rate ECRS design.

## 7. Discussion

The challenge of the post-lockdown economy will be to balance protection and reallocation of workers. During COVID-19, many liberal welfare states introduced new temporary wage subsidy schemes (Eurofound, 2020). As sanitary and economic conditions improve, temporary wage subsidy schemes will be phased out or integrated with the pre-existing system. As emergency schemes are reformed, policy makers face two challenges. First, successfully reforming emergency schemes requires appropriate timing. Timing reform will be particularly challenging given the uncertainty around the duration of the crisis and the state of the post-pandemic economy. With sectors and occupations recovering at different speeds, phasing out job retention schemes too early will cause loss of viable businesses and excessive unemployment, and may not be political feasible. Retaining schemes for too long supports unviable firms, prevents workers from relocating to higher productivity jobs and maintains poor work incentives. Second, policy makers will decide to discontinue the scheme or to integrate it with the existing tax-benefit system. Policy makers learned during the crisis and will draw lessons for future emergency responses. The swift introduction of new schemes to support middle and high earners required reform and generate transaction costs. Fortifying existing automatic stabilizers could avoid frequent reform and transaction costs and reduce uncertainty for businesses and workers in future crisis.

The Irish experience illustrates that novel schemes often require reform and timing of reform can determine the designs viability and political feasibility. During COVID-19, subsequent reforms of the Irish wage subsidy addressed different issues. Initially, the breadth of the shock and the large emphasize on solidarity meant that concerns over deadweight losses and substitution effects were secondary (OECD, 2020). The initial response was crude, and subsequent reforms increased payment levels and altered the payment structure, making the scheme more targeted. In Ireland, the introduction of generous, earnings-related pro-rata payments presented a departure from the traditional system, which predominantly relied on flat-rate or means-tested payments. This led to a mismatch in in-work and out-of-work support and poor work incentives. Poor work incentives, particularly at for low earners, resulted from partial earnings replacement under the wage subsidy and a generous unemployment payment. Reforms during the early months of crisis addresses incentive issues. Later, when economic conditions improved and the emphasize shifted towards reintegration, reforms addressed concerns over fiscal sustainability, deadweight losses and widespread fears that poor incentives to work would worsen medium run labour market outcomes (OECD, 2021). The effort to reintegrate the scheme and reduce the generosity of support payments however only shortly preceded a second wave of COVID-19. During a second wave, policy makers receded on their efforts to reduce generosity and reintegrate support schemes. Poor timing of reform meant that the scheme was reformed again.

Reforms responded to emerging issues but generated transaction costs and uncertainty for firms (OCED 2020a). Reforms require firms to invest efforts in understanding the scheme and training employees, increasing the administrative and compliance cost of the scheme (Bennett et al, 2009). Administrative and compliance costs grow with the complexity of the scheme and the frequency of design changes. CWS reforms substantially altered the structure of payments. Early reforms made the scheme more targeted and increased its complexity, increasing transaction costs [7]. Whilst costly, reforms addressed issues with the scheme's functioning, and policy makers learned about designs' viability and political acceptability. The last reform proposed a simpler design which improved work incentives by matching CWS' and PUP's payment structure. Frequent reforms thus increased transaction costs but simultaneously improved the economic and political viability of the scheme.

Like any policy action, policy makers' response to emergencies is contingent on the political and social context (Hale et al, 2020). Some commentators propose that the generous policy response to COVID-19 results from Ireland's risk averse and consensus oriented political culture (Hick and Murphy, 2021). Nolan (2020) suggest that the ruling party Fine Gael was interested in supporting their middle-class voters, many of whom were without work for the first time. Figure 3 provides some support to this claim, showing the COVID-19 government policy stringency index (Hale et al, 2021) against the average and median compensation rate. CWS became more generous as the level of restrictions and the number of those affected rose, particularly to middle earners. The September reform, which was announced as part of a new stimulus package in July when restrictions were low, reduce the level of income replacement for the first time. Policy makers however swiftly raised payment levels of higher incomes when restrictions were increased and demand for the scheme rose, suggesting that the level of earnings insurance offered by the scheme was responsive to demand for the scheme. Indeed, there is concern that emergency schemes, if the timing of phase out and the level of the subsidy is not clarified, are open to political abuse (OECD, 2020).

---

[7] Transaction costs are here estimated qualitatively from Table 5 in the appendix.

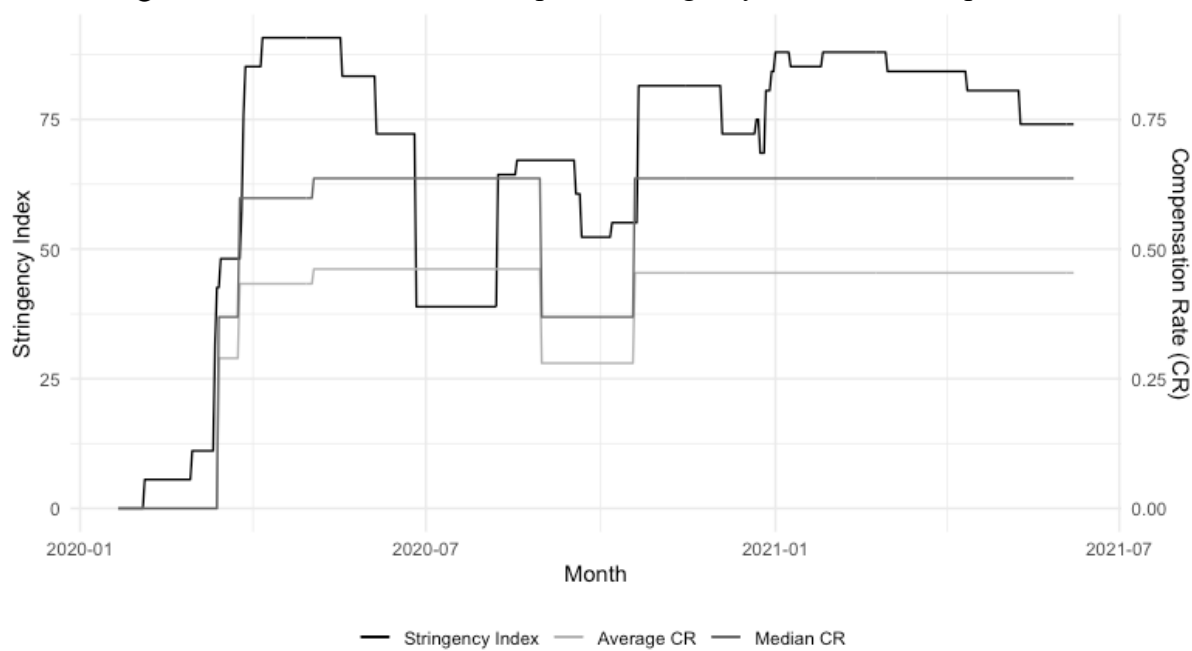

Figure 3. Ireland COVID-19 response Stringency Index and Compensation rates

Note: Data for the Stringency index is provided by the Oxford COVID-19 Government response tracker (OxCGRT) (Hale et al., 2021). Downloaded from https://www.bsg.ox.ac.uk/research/research-projects/covid-19-government-response-tracker on June 15th 2021.

Understanding whom resources were directed to becomes increasingly important with rising pressures to reduce spending and public debt. During the COVID-19 crisis, fiscal spending was financed through debt and countries accumulated significant public debt (O'Donoghue et al, 2021). Public spending was necessary to prevent the short but sharp recession turning into a long depression (Saez and Zucman, 2020). As restrictions are lifted and economies open, policy makers will look to reduce emergency spending and public debt. In Ireland, Minister of Public Expenditure and Reform Michael McGrath announced that spending will be tightly controlled in the Government's bid to reduce public spending by €12 billion in 2022 (McConnell, 2021). Spending cuts introduce equity concerns, as lower earners are likely impacted most heavily. Overall, the tax-benefit system supported low earners that lost their job due to COVID-19 most, but subsequent reforms reduced PUP payments for lower earners. For those remaining at work, our analysis shows that wage subsidy payments were regressive, paying more at higher earnings. Particularly low-wage earners supported by the wage subsidy may be net losers in the medium run.

Over the coming months, the emphasize of temporary wage subsidies will shift away from protecting all jobs to supporting jobs in vulnerable employment only (OECD, 2020). Emergency wage subsidies could be repurposed as temporary hiring subsidies (OECD, 2021). Evidence from the Great Recession suggests that hiring subsidies can effectively boost employment (Cahuc et al, 2018). Blanchard et al (2020) suggest a phasing out of emergency wage subsidies and a phasing in of sectorial subsidies. A marginal wage subsidy, which pays for additional employment only, could assist in reallocating workers to expanding sectors. Sectorial subsidies could support sectors subject to legal restrictions, while a marginal wage subsidy could increase employment at moderate fiscal cost. Alternatively, policy makers could more heavily rely on existing short time working (STW) schemes. In liberal welfare states, STW schemes are in line with flat-rate unemployment payments. STW schemes increase workers' incentive look for alternative employment and eliminate firms' incentive to restrict turnover. STW does however not allow firms to continue production whilst helping them meet their liquidity needs. Withdrawal of the wage subsidy could thus cause viable businesses in sectors still affected by the crisis to close.

A key policy question pertains whether liberal welfare states can protect middle- and higher-income households in future crisis. In the future, questions of timing, transaction costs, political motives and equity concerns may be reduced by permanently integrating emergency schemes as automatic stabilizers. During the COVID-19 crisis, policy makers learned and improved the functioning of the scheme through successive reforms. The last design matched in-and-out-of-work payments, improved work incentives and reduced program costs by simplifying the scheme's design whilst providing high levels of income replacement. Reforms leading up to this design were however frequent and costly. The need for costly reform and the scope for politically motivated design can be reduced by extending build-in automatic stabilizers to protect middle-earning and high-earning jobs. Access to enhanced support schemes could be contingent on a sectorial turnover rule or the level of legal restrictions (OCED, 2020a). Such rules help ensure the timeliness and predictability of support, but also limit the fiscal burden through improved targeting and reduced need for reform.

## 8. Conclusions

In this paper, we reviewed the structure and functioning of a novel emergency wage subsidy scheme introduced to a liberal welfare state during COVID-19. The scheme operated under multiple designs. We compare how different designs affect the level of income replacement, work incentives and income inequality. By our knowledge, this is the first study of emergency wage subsidy comparing the functioning of an emergency wage subsidy introduced to a liberal welfare scheme under different designs. This study employs a microsimulation-nowcasting approach to provide insights into the impact of wage subsidy's design on work incentives and the income distribution, informing policy makers on the challenges faced in designing future emergency response tools. We consider the specificities of the COVID-19 crisis by proposing an augmented disposable income concept, which includes changes in work-related expenditures and capital returns.

We find that all wage subsidy designs were most generous to those at the bottom of the earnings distribution. Flat-rate designs protect lower earning jobs relatively more and pro-rata designs direct resources towards middle earners. Designs became more generous towards the middle of the distribution as government restrictions and demand for the scheme grew. All designs provided a substantial number of workers with strong work disincentives. Matching the structure of out-of-work and in-work support reduces work disincentives, but strong work disincentives remain for a substantial share of workers. We found large increases in labour market inequality during the emergency, but the policy response largely cushioned the impact on disposable income inequality. Once taxes and other benefits are considered, inequality in disposable income was lowest under generous designs. Generous schemes reduce the increase in disposable income inequality most, irrespective of their structure.

The swift introduction of novel emergency schemes to protect jobs and support incomes led to poor work incentives. Reforms addressed incentive issues and adjusted the level of income replacement. Frequent reform however led to high transaction costs and uncertainty. It appears that reforms were in part motivated by political considerations, so that workers were better protected when demand for the scheme was highest. Increases in subsidy payment levels were particularly beneficial for workers situated in the middle of the earnings distribution. The introduction of a pro-rata wage subsidy and multi-tier enhanced unemployment payments marked a departure from the existing tax-benefit system. As sanitary and economic conditions improve, emergency schemes will be reintegrated or phased-out. We discussed issues around timing and nature of reintegration. We suggest a phasing out of emergency wage subsidies and a phasing in of sectorial subsidies contingent on legal restrictions. Additionally, a marginal wage subsidy could be implemented to support expanding industries and help reallocate workers that lost their job during COVID-19. To prepare for future crisis, we suggest that introducing automatic stabilizers that support middle and high earners could reduce the need for costly reform, uncertainty, and the scope for political abuse. Automatic stabilizers could be implemented as wage subsidies contingent on sectorial turnover. Stress-testing the tax-benefit system using ex-ante methods could help design automatic stabilizers and avoid costly reform in the future.

# 10. Appendix

*Appendix 1: Description of the Emergency Wage Subsidy Scheme's structure*

Table 5.    Changes in the Emergency Wage Subsidy Scheme Design

| Date | Program | Type | Earnings | Payment Level |
|---|---|---|---|---|
| 15 March | ECRS | Flat rate | All | €203 |
| 26 March | TWSS transitional phase | Tiered approach | €0 to €586 APNP | 70% max €410 |
| | | | €586 to €960 APNP | 70% max €350 |
| | | | €960 APNP | €0 |
| May 4th | TWSS operational phase | Tiered approach | €0 to €412 APNP | 85% previous net pay |
| | | | €412 to €500 APNP | €350 |
| | | | €500 to €586 APNP | 70% previous net pay |
| | | | €586 to €960 APNP | Tapered approach |
| | | | >€960 | Tapered approach |
| September 1st | EWSS | Tiered approach | <€151.50 GP | €0 |
| | | | €151.50 to €202.99 GP | €151.50 |
| | | | €203 to €1462 GP | €203 |
| | | | >€1462 GP | €0 |
| October 20th | EWSS 2 | Tiered approach | <€151.50 GP | €0 |
| | | | €151.50 to €202.99 GP | €203 |
| | | | €203 to €299.99 GP | €250 |
| | | | €300 to €399.99 GP | €300 |
| | | | €400 to €1462 GP | €350 |
| | | | >€1462 GP | €0 |

Note: APNP – Average Previous Net Pay; GP – Gross Pay

*Appendix 2: Description of Pandemic Unemployment Payment's structure*

Table 6.    Changes in Pandemic Payment over time

| Date | Pre 24-Mar | 24-Mar | 29-Jun | 17- Sep | 16-Oct |
|---|---|---|---|---|---|
| Previous Earnings | Payment Level | | | | |
| +400 | 203 | 350 | 350 | 300 | 350 |
| 300-399.99 | 203 | 350 | 350 | 300 | 300 |
| 200-299.99 | 203 | 350 | 350 | 250 | 250 |
| <200 | 203 | 350 | 203 | 203 | 203 |

*Appendix 3 : CWS and PUP Budget constraint*

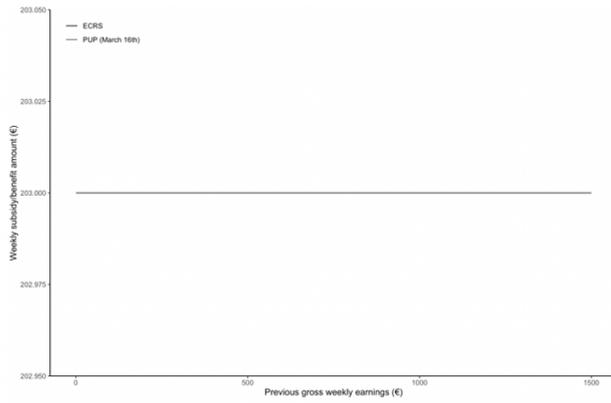
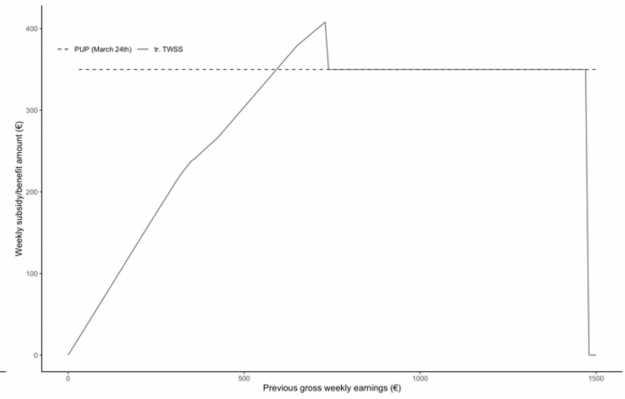
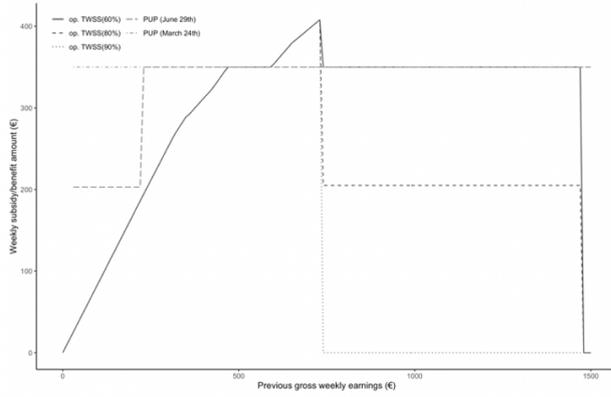
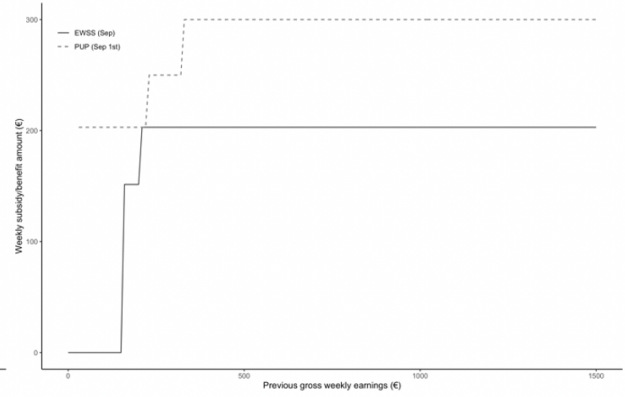
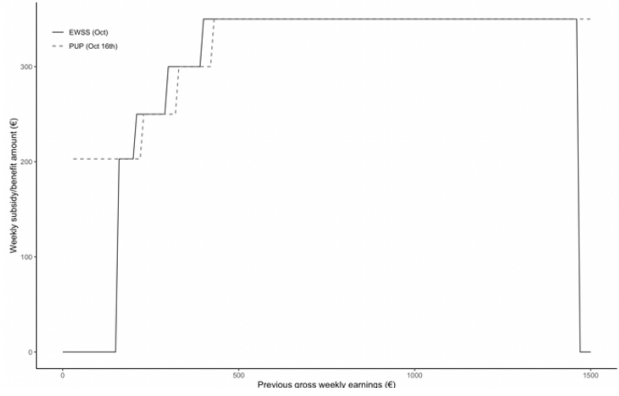

*Appendix 4 : Table with administrative and compliance costs*

| Scheme | Institutional structure | Number of payment bands | Administrative steps | Information requirements | Reimbursement frequency | Conditionality |
|---|---|---|---|---|---|---|
| ECRS | single flat rate | 1 | (1) register at ROS (2) Submit Self-Declaration; (3) set PRSI class to J9; (4) enter non-taxable amount of €203; (5) set gross pay at €0.01 | (1) Number of employees temporarily laid off | Weekly | (1) employees on payroll in February/March 2020 temporarily laid off; (2) employer is up-to-date with their payroll submission; (3) retain workers on payroll |
| TWSS (transitional) | earnings related pro-rata | 2 | (1) register at ROS; (2) Submit Self-Declaration; (3) set PRSI class to J9; (4) enter non-taxable amount (worker ANWP) | (1) Declare eligible employees; (2) Eligible workers' ARNWP; (3) Subsidy received + additional payments by employer; (4) maintain data on qualification criteria | Weekly | (1) reduction of 25% of turnover (2) employees on payroll in February/March 2020 temporarily laid off; (3) retain workers on payroll |
| TWSS (operational) | earnings related flat rate and pro-rata mix, tapered approach for top incomes | 5 | (1) register at ROS; (2) Submit Self-Declaration; (3) set PRSI class to J9; (4) enter non-taxable amount (worker ANWP) | (1) Declare eligible employees; (2) Eligible workers' ARNWP; (3) Subsidy received + additional payments by employer | Weekly | (1) reduction of 25% of turnover; (2) employees on payroll in February/March 2020; (3) retain workers on payroll |
| EWSS | Earnings related flat rate | 4 | (1) register at ROS; (2) Submit Self-Declaration; (3) set PRSI class to J9; (4) enter non-taxable amount (worker ANWP) | (1) Declare eligible employee; (2) employee GP; (3) information on compliance checks + monthly eligibility review; (4) information on directors, shareholders, partnerships | Monthly (adjusted to weekly) | (1) reduction of 30% of turnover from July-December 2020; (2) employees on payroll in February/March 2020 (3) retain workers on payroll |
| EWSS | Earnings related flat rate | 6 | (1) register at ROS; (2) Submit Self-Declaration; (3) enter €0.00 or €1.00 as 'Other payment'; (4) Submit normal PSRI amount | (1) number of eligible employees; (2) employee GP; (3) information on compliance checks + monthly eligibility review; (4) information on directors, shareholders, partnerships. | Monthly (adjusted to weekly) | (1) reduction of 30% of turnover from July-December 2020; (2) employees on payroll in February/March 2020 1 temporarily laid off; (3) retain workers on payroll |

*Appendix 5 : CWS cost per recipient*

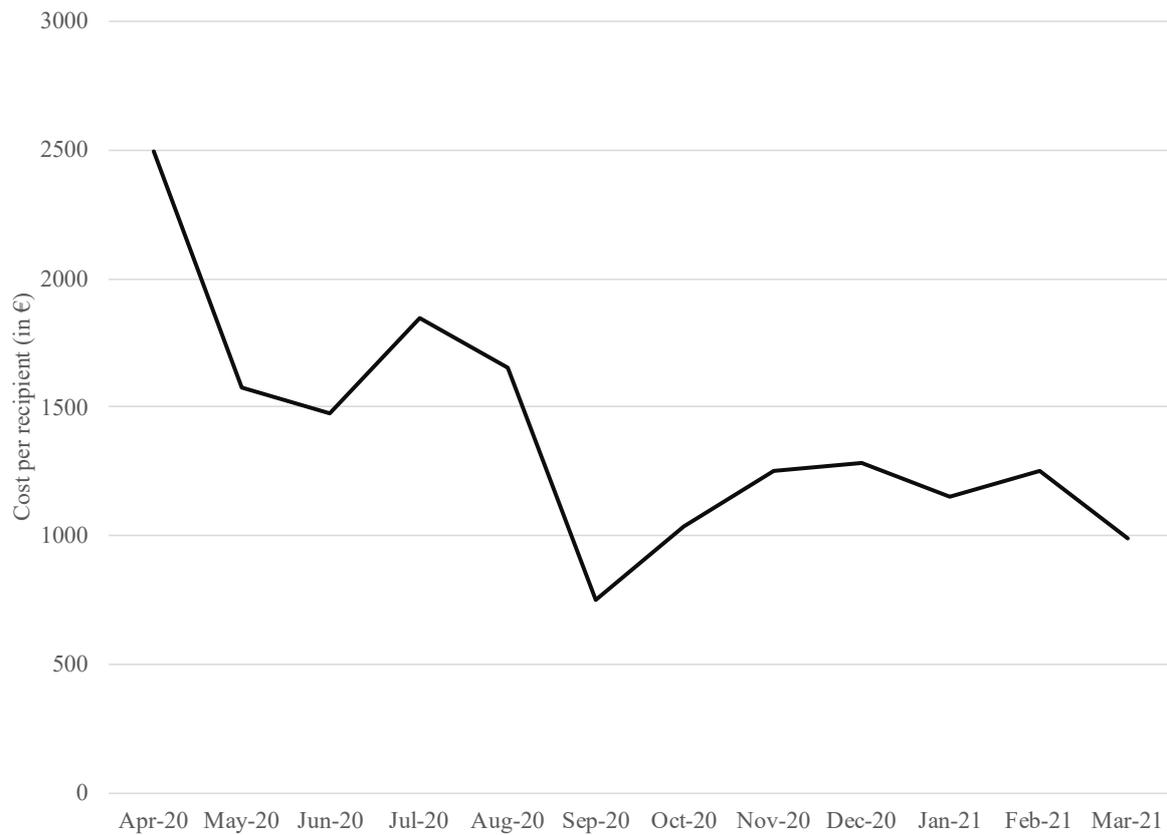

Source : https://www.revenue.ie/en/corporate/information-about-revenue/statistics/number-of-taxpayers-and-returns/covid-19-support-schemes-statistics.aspx

*Appendix 6 : CWS throughout the three waves of COVID-19 in Ireland*

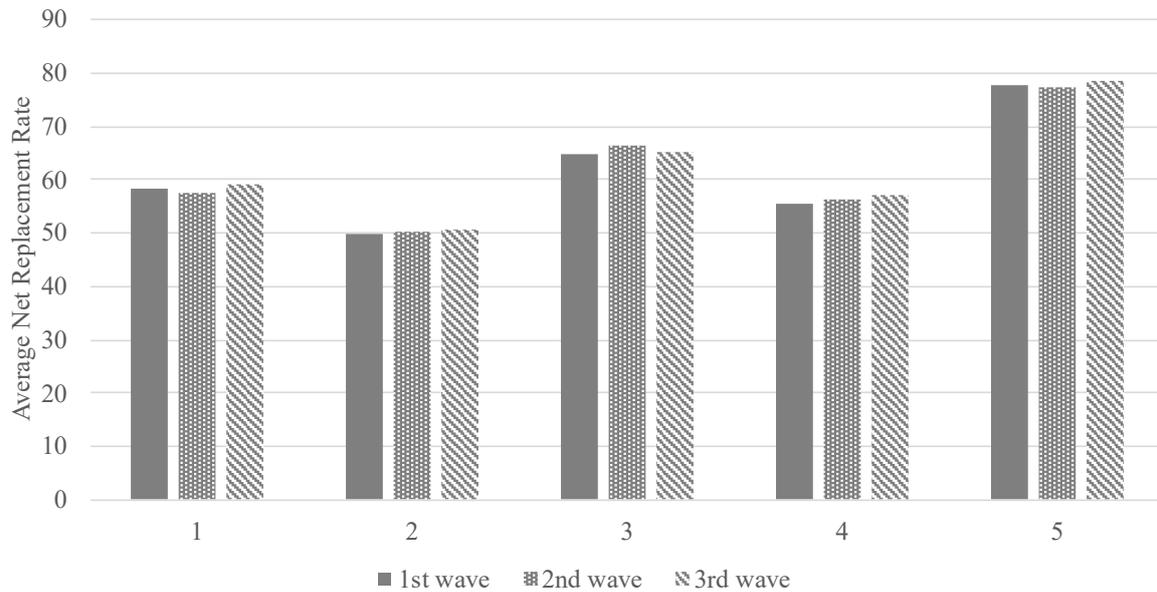

Figure 4. Average net replacement rate by design by COVID-19 wave

Note : 1= ECRS; 2= tr. TWSS; 3= op. TWSS; 4= EWSS (Sep); 5= EWSS (Oct); Wave 1 = May 2020; Wave 2 = November 2020; Wave 3 = January 2021 ; Net replacement rates are computed based on household adjusted disposable income equivalized using the LIS equivalent scale (household income divided by square root of household size).

Figure 5. Share of recipients with relative replacement rate ($RR_{relative}$) larger than 1 by COVID-19 wave

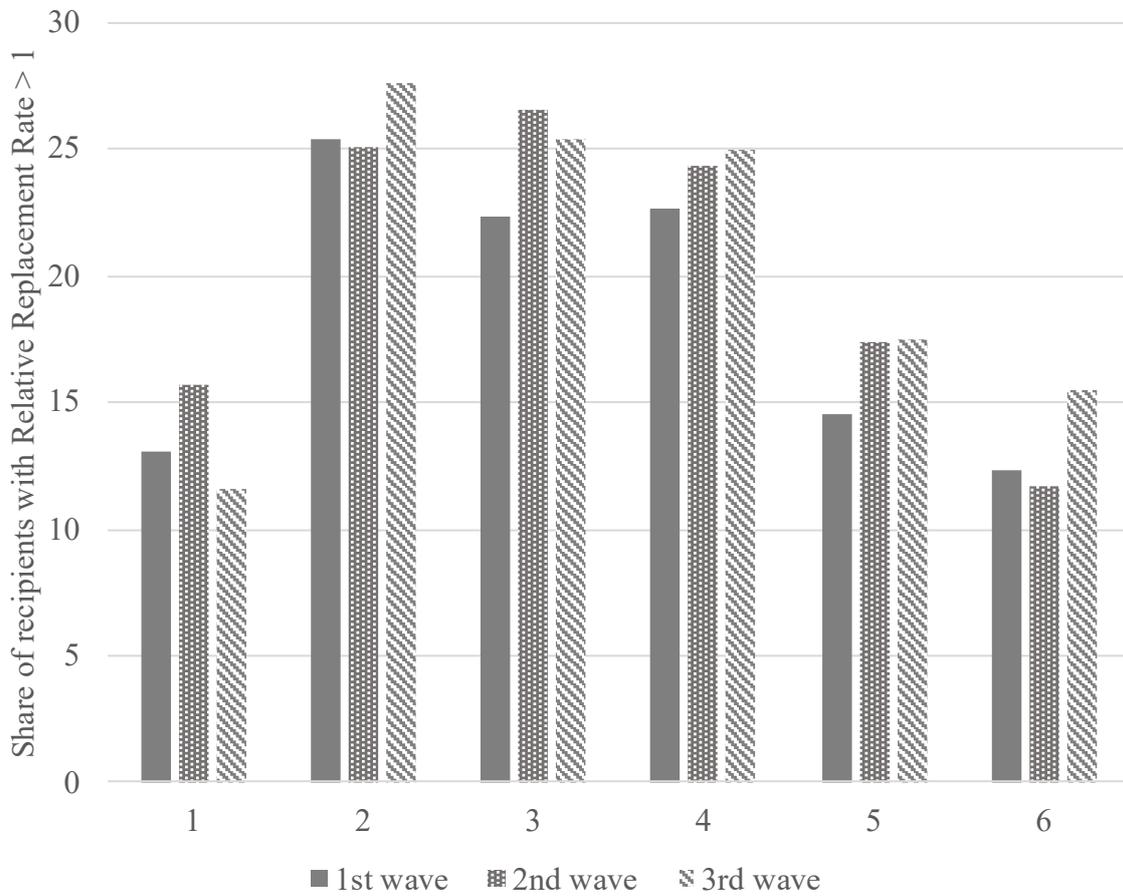

Note : 1= 15-Mar (Flat-Rate CWS and Low Flat-Rate PUP); 2= 26-Mar (Earnings-Rate CWS and High Flat-Rate PUP); 3= 4-May (Earnings-Rate CWS and High Flat-Rate PUP); 4= 1-Jul (Earnings-Rate CWS and Two-Rate PUP); 5= 1-Sep (Flat-Rate CWS and Three-Rate PUP) ; 6= 1-Oct (Flat-Rate CWS and Four-Rate PUP)

Table 7. Table: Share of recipients with relative replacement rate ($RR_{relative}$) higher than 1 and 0.7

| $RR_{relative}$ (rRR) | ECRS | tr. TWSS | op. TWSS (May) | op. TWSS (July) | EWSS (Sep) | EWSS (Oct) |
|---|---|---|---|---|---|---|
| 1st wave | | | | | | |
| rRR > 1 | 13.0 | 25.4 | 22.4 | 22.7 | 14.6 | 12.3 |
| rRR > 0.7 | 47.2 | 72.3 | 71.4 | 71.0 | 56.6 | 62.3 |
| 2nd wave | | | | | | |
| rRR > 1 | 15.7 | 25.0 | 26.5 | 24.3 | 17.4 | 11.7 |
| rRR > 0.7 | 54.6 | 71.7 | 73.4 | 73.7 | 65.9 | 66.2 |
| 3rd wave | | | | | | |
| rRR > 1 | 11.6 | 27.6 | 25.3 | 25.0 | 17.5 | 15.5 |
| rRR > 0.7 | 50.1 | 78.0 | 75.7 | 78.4 | 65.5 | 66.4 |

Note: 15-Mar = Flat-Rate CWS and Low Flat-Rate PUP; 26-Mar = Earnings-Rate CWS and High Flat-Rate PUP; 4-May = Earnings-Rate CWS and High Flat-Rate PUP; 1-Jul = Earnings-Rate CWS and Two-Rate PUP; 1-Sep = Flat-Rate CWS and Three-Rate PUP ; 1-Oct = Flat-Rate CWS and Four-Rate PUP

Table 8. Progressivity and redistribution of CWS on equivalized household disposable income

| Gini | ECRS | tr. TWSS | op. TWSS (May) | op. TWSS (July) | EWSS (Sep) | EWSS (Oct) |
|---|---|---|---|---|---|---|
| (1) Gini in Market income (incl. CWS) | | | | | | |
| 1st wave | 59.2 | 60.0 | 60.8 | 61.6 | 59.4 | 60.3 |
| 2nd wave | 57.0 | 56.8 | 57.7 | 58.3 | 56.5 | 57.6 |
| 3rd wave | 57.1 | 56.9 | 57.6 | 58.6 | 56.2 | 58.9 |
| (2) Gini in Gross income ((1) + benefits, incl. PUP) | | | | | | |
| 1st wave | 36.9 | 35.8 | 35.3 | 35.7 | 36.4 | 35.5 |
| 2nd wave | 37.4 | 36.3 | 36.3 | 36.6 | 36.7 | 36.0 |
| 3rd wave | 37.5 | 36.6 | 36.1 | 36.6 | 36.3 | 36.8 |
| (3) Gini in Adjusted disposable income ((2) – taxes – work related costs) | | | | | | |
| 1st wave | 31.8 | 31.1 | 30.8 | 30.9 | 31.9 | 31.0 |
| 2nd wave | 32.4 | 31.6 | 31.5 | 31.6 | 31.9 | 31.4 |
| 3rd wave | 3 2.2 | 31.6 | 31.3 | 31.7 | 31.5 | 32.0 |
| (4) Gini in Adjusted disposable income without CWS ((3) – CWS) | | | | | | |
| 1st wave | 34.3 | 33.2 | 34.2 | 35.1 | 34.0 | 34.8 |
| 2nd wave | 35.1 | 33.8 | 34.9 | 35.7 | 34.3 | 35.4 |
| 3rd wave | 34.8 | 33.5 | 34.5 | 35.8 | 33.9 | 36.5 |
| (5) Benefit redistribution (RS) ((4) – (3)) | | | | | | |
| 1st wave | 2.5 | 2.1 | 3.4 | 4.2 | 2.1 | 3.8 |
| 2nd wave | 2.7 | 2.2 | 3.4 | 4.1 | 2.4 | 4.0 |
| 3rd wave | 2.6 | 1.9 | 3.2 | 4.1 | 2.4 | 4.5 |
| (6) Benefit Regressivity (K) | | | | | | |
| 1st wave | 0.63 | 0.65 | 0.65 | 0.64 | 0.66 | 0.67 |
| 2nd wave | 0.61 | 0.65 | 0.65 | 0.64 | 0.66 | 0.64 |
| 3rd wave | 0.64 | 0.67 | 0.67 | 0.64 | 0.67 | 0.63 |

Note : Computed based on a sample of households with one member in receipt of CWS only. Computations based on household adjusted disposable income equivalized using the squared adult equivalent scale. Variation in market income reflect differences in simulation results; K = Kakwani index; RS = Reynolds Smolensky index.